\documentclass{aa}
\usepackage[varg]{txfonts}
\usepackage{natbib,twoopt}
\usepackage[breaklinks=true]{hyperref} 
\bibpunct{(}{)}{;}{a}{}{,} 
\usepackage{graphicx}
\usepackage{float}
\usepackage{color, colortbl}
\usepackage{verbatim}
\usepackage[normalem]{ulem}
\definecolor{LightCyan}{rgb}{0.88,1,1}
\usepackage{amsmath}

\begin{document}

\title{The X-ray view of the Hyades cluster: updated}
\author{S. Freund \and J. Robrade \and P.C. Schneider \and J.H.M.M. Schmitt}
\institute{Hamburger Sternwarte, Universit\"at Hamburg, 21029 Hamburg, Germany \\
e-mail: sebastian.freund@uni-hamburg.de}
\abstract{}
{We revisit the X-ray properties of the main-sequence Hyades members and the relation between X-ray emission
and stellar rotation.}
{As input catalog for Hyades members, we combined three recent Hyades membership lists derived from \textit{Gaia} 
DR2~data including the Hyades core and its tidal tails. We searched for X-ray detections from 
the \textit{ROSAT} all-sky survey (RASS) and pointings from \textit{ROSAT}, the \textit{Chandra} X-Ray 
Observatory, and \textit{XMM-Newton} of the main-sequence Hyades members. Furthermore, we adopted 
rotation periods derived from \textit{Kepler}'s K2 mission and other resources.}
{We find an X-ray detection for 281 of 1066 bona fide main-sequence Hyades members and provide statistical upper 
limits for the undetected sources. The majority of the X-ray detected stars is located in the Hyades core 
because of its generally smaller distance to the Sun. F- and G-type stars have the highest detection 
fraction (72~\%), while K- and M-type dwarfs have lower detection rates (22~\%). The X-ray luminosities of 
the detected members range from $\sim 2\times 10^{27}$~erg~s$^{-1}$ for late M-type dwarfs to 
$\sim 2\times 10^{30}$~erg~s$^{-1}$ for active binaries. 
The X-ray luminosity distribution functions formally differ for the members in the core and tidal tails, which is likely caused by a larger fraction of field stars in our Hyades tails sample. Compared to previous studies, our sample is slightly fainter in X-rays due to differences in the used Hyades membership list, furthermore, we extend the X-ray luminosity distribution to fainter luminosities. The X-ray activity of F- and G-type stars is well defined at 
$F_X/F_\mathrm{bol} \approx 10^{-5}$. The fractional X-ray luminosity and its spread increases to later 
spectral types reaching the saturation limit ($F_X/F_\mathrm{bol} \approx 10^{-3}$) for members later than 
spectral type M3. Confirming previous results, the X-ray flux varies by less than a factor of three between epochs for the 
104 Hyades members with multiple epoch data, significantly less than expected from solar-like activity 
cycles. 
Rotation periods are found for 204 Hyades members, with about half of them being detected in X-rays. 
The activity-rotation-relation derived for the coeval Hyades members has properties very similar to 
those obtained by other authors investigating stars of different ages.}{}
\maketitle

\section{Introduction}
\label{sec: introduction}
Open star clusters provide ideal laboratories to study the activity-rotation-age relation of late-type 
stars because all their members are coeval. As the nearest well-populated open cluster, the Hyades 
with a distance of only $\sim 47$~pc to the Sun and an age of about 640~Myr \citep{lodieu19} is of 
unique importance. The second \textit{Gaia} Data Release \citep[\textit{Gaia} DR2]{GaiaDR2,GaiaMission}, containing highly 
accurate parallaxes and proper motions for 1.3 billion sources and radial velocities for about 
7 million sources, significantly improves also the membership identification of the Hyades. 
Adopting these data, \citet{lodieu19} present a revised census of the members of the Hyades core. 
According to \citep{roeser11}, Hyades members within a distance of 9~pc are gravitationally bound, 
but for stars at larger distances, the Galaxy exerts tidal forces leading to the creation of tidal 
stellar tails, which have been discovered independently by \citet{meingast19} and \citet{roeser19} 
using data from \textit{Gaia} DR2.

The investigation of rotation periods for Hyades members has a long history and the Hyades are among
the first open clusters for which photometric rotation periods were measured for low-mass 
stars \citep{radick87,radick95}. Since then, the number and quality of available rotation periods has increased substantially especially due to \textit{Kepler}'s K2 mission \citep{K2mission}. With 
its $\sim 100$~deg$^2$ field of view, \textit{Kepler} targeted two slightly different parts of 
the Hyades in K2 Campaign 4 and 13 for about 75 days each. \citet{douglas16,douglas19} use these 
K2 data to derive new rotation periods for Hyades members (for more details see Sect.~\ref{sec: Rotation periods}). 

X-ray emission from the Hyades members has been systematically investigated ever since the 
\textit{Einstein Observatory} era \citep{stern81,micela88}. As Hyades input catalog, \citet{micela88} combined 323 certain or probable Hyades members from different optical catalogs, arguing that their catalog is complete down to the 
9$^{th}$ magnitude. 66 of 121 Hyades members covered by a pointing of the \textit{Einstein Observatory} are detected,
and depending on the individual exposure time and the off-axis angle of the considered Hyades,
\citet{micela88} reach detection limits of $\sim 2 - 7\times 10^{28}$~erg~s$^{-1}$ at the assumed Hyades distance of 45~pc. The first complete X-ray survey of the Hyades cluster region was performed by \citet{stern95} using data from the \textit{ROSAT} all-sky survey (RASS). They discuss the X-ray properties of 440 optically selected Hyades members including fainter sources than the input catalog of \citet{micela88}. With an estimated detection limit of $\sim 1.5 - 3\times 10^{28}$~erg~s$^{-1}$ at 45~pc, \citet{stern95} detect 187 of their Hyades members as X-ray sources and derive upper limits for the undetected sources. 
\textit{ROSAT} PSPC pointed observation of the 
Taurus-Auriga-Perseus region including the Hyades are analyzed by \citet{stelzer01}. Now, new data 
reductions for the RASS \citep{boller16} and \textit{ROSAT} PSPC pointed observations \citep{2RXP-catalog} are available, and furthermore, observations with \textit{XMM-Newton} and the \textit{Chandra} X-Ray Observatory \citep{weisskopf02} provide new Hyades detections with unprecedented sensitivity and accuracy allowing to detect sources with later spectral types and lower activity. 

In this paper, we therefore revisit the X-ray properties and the activity-rotation relation of the Hyades cluster 
applying current membership lists including the tidal tails, rotation periods, and significantly improved 
X-ray detection lists. We discuss the data acquisition in Sect.~\ref{sec: Data acquisition} describing the 
membership list of the Hyades in Sect.~\ref{sec: Membership}, the rotation periods in 
Sect.~\ref{sec: Rotation periods}, and the X-ray observations from the different instruments 
in Sect.~\ref{sec: X-ray observations}. In Sect.~\ref{sec: Properties of the Hyades} we present 
our results discussing the X-ray detection rates in Sect.~\ref{sec: x-ray detection rates}, the 
distribution of the X-ray luminosity and activity in Sect.~\ref{sec: X-ray activity and HRD}, the 
variability of the Hyades members in Sect.~\ref{sec: Variability}, and the activity-rotation 
relation in Sect.~\ref{sec: Rotation properties}. Finally, we draw our conclusions in 
Sect.~\ref{sec: Conclusion}.

\section{Data acquisition}
\label{sec: Data acquisition}
\subsection{Membership}
\label{sec: Membership}

We obtained our new Hyades member sample from three recent publications, which identify
Hyades members based on data from \textit{Gaia} DR2. \citet{lodieu19} studied the central region of the Hyades and associate 710 Gaia sources with the Hyades, located within 30~pc to the cluster center; they estimate the contamination of this sample to be around $ 5 - 10~\%$.

The tidal tails of the Hyades were independently discovered by \citet{meingast19} and \citet{roeser19}. Both publications started from the same data, i.e. the \textit{Gaia} DR2 sources within a 200~pc sphere around the Sun, but different selection criteria were adopted to extract the Hyades' core and tidal tail members. The most important difference is that \citet{roeser19} identified the Hyades members by applying the convergent-point method solely based on tangential velocities, while \citet{meingast19} relied on 3D space velocities, excluding all sources without a radial velocity measured in \textit{Gaia} DR2. However, radial velocities are only available for \textit{Gaia} DR2 sources with G magnitudes between 4 and 13~mag and effective temperatures in the range of 3550 to 6900~K. \citet{roeser19} subdivide their sample in the core containing sources within 18~pc around the cluster center and the leading and trailing tail selected by eye.

For the following analysis, we adopted the 979 sources from \citet{roeser19} classified as core or tail sources and not flagged as possible interlopers. According to \citet{roeser19}, $\sim 1.4~\%$ and $\sim 13~\%$ of the sources in the core and in the tails are spuriously associated to the Hyades. Furthermore, we obtained 238 Hyades members from \citet{meingast19}, who do not comment on the reliability of their sample.   

\begin{figure}[t]
	\resizebox{\hsize}{!}{\includegraphics{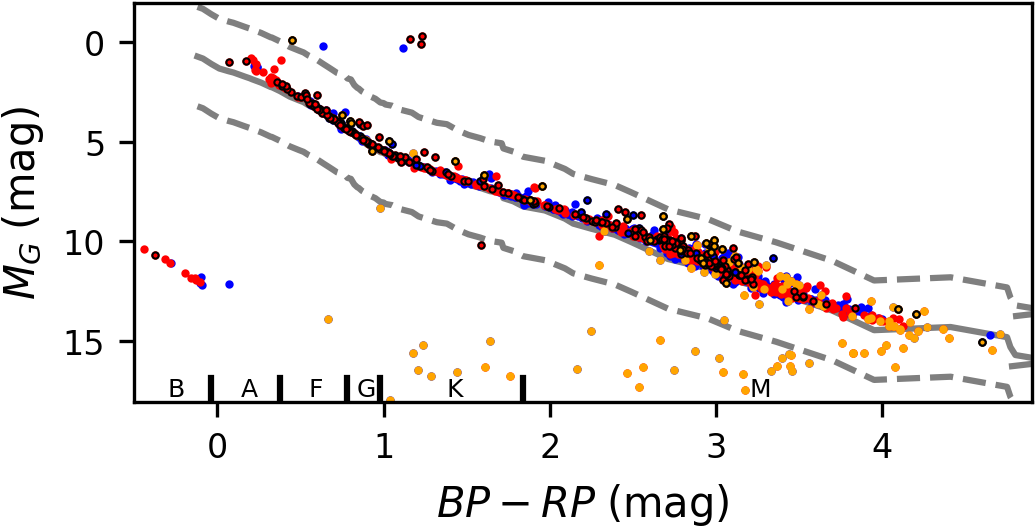}}
	\caption{Hertzsprung-Russell diagram (HRD) of the Hyades. The red and blue sources indicate 
Hyades members in the core 
and in the tails, respectively. The orange sources are only associated to the Hyades by \citet{lodieu19}, 
sources outlined in black are detected X-rays. The theoretical main-sequence adopted 
from \citet{color-table} is shown by the black solid line, while the dashed lines indicate the values 2.5~mag brighter and fainter than the main-sequence.}
	\label{fig: Hyades member HRD}
\end{figure}

Combining the three Hyades catalogs, we obtained 1142 unique bona fide members. The Hertzsprung-Russell diagram (HRD) of the so selected
Hyades members (Fig.~\ref{fig: Hyades 
member HRD}) shows that most of the sources are located on the main-sequence. The four Hyades giants 
are located well above the main-sequence \citep[and discussed in ][]{schroeder19}
and the white dwarfs are found at $M_\mathrm{G} < 10$~mag 
and $BP-RP < 0$~mag in the HRD. Furthermore, there are many sources that are also clearly fainter than expected for main-sequence stars but have red colors. We inspected these sources and found that they are only 
associated to the Hyades by \citet{lodieu19} and are excluded by \citet{roeser19} because they do not meet the quality criteria of a small ``unit weight error'' and ``flux excess ratio'' proposed by \citet{lindegren18} and \citet{babusiaux18}. Therefore, the positions of these sources in Fig.~\ref{fig: Hyades member HRD} are probably wrong. In the following analysis, we therefore concentrate on main-sequence stars excluding all sources that are 2.5~mag brighter or fainter than expected for dwarfs, thus obtaining our sample of 1066 bona fide 
Hyades members.
This brightness cut  excludes most of the rather peculiar red sources below 
the main-sequence (cf., Fig.~\ref{fig: Hyades 
member HRD}). We expect that the remaining sources are highly probable Hyades members thanks 
to their location on or very close to the Hyades main-sequence 
although some do not meet the formal ``unit weight error'' and ``flux excess ratio'' criteria.

\begin{figure}[t]
\resizebox{\hsize}{!}{\includegraphics{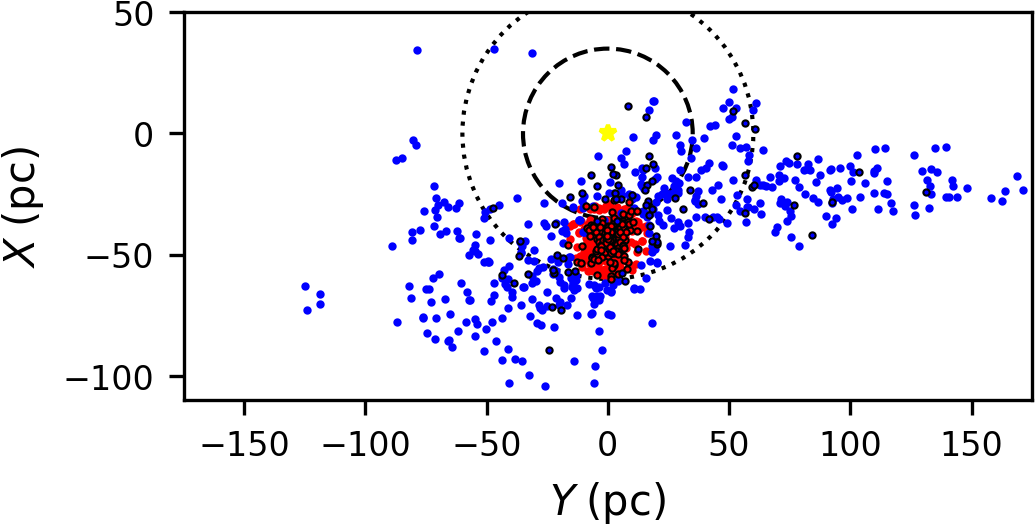}}
\caption{Hyades members in Galactic Cartesian XY coordinates. The red and blue sources are associated with the cluster's core and tidal tails, respectively. Sources outlined in black are detected in X-rays. The star marker indicates the position of the sun and the dashed and dotted line show 35~pc and 60~pc radius around the sun, respectively.}
\label{fig: Hyades member classification}
\end{figure}

In Table~\ref{tab: Catalogs of the Hyades} we denote the catalogs providing the Hyades members. 
To distinguish between core and tidal tails, we applied the condition of \citet{roeser19} and classify Hyades member within 18~pc to the cluster center as core sources and members with larger distances as tidal tail sources adopting the position of the cluster center from \citet{lodieu19}. Thus, finally 550 Hyades are associated with the 
core and 516 with the tidal tails. In Fig.~\ref{fig: Hyades member classification} we show the spatial 
distribution in Galactic Cartesian coordinates of the Hyades members discussed in this paper, where the 
X- and Y-axes are directed to the Galactic center and in the direction of the Galactic rotation, respectively. 
Most core sources have similar distances, some tail members are in the immediate solar vicinity, most tail 
members, however, have much larger distances than the Hyades core.
\begin{table}[t]
	\centering%
	\caption{Catalogs of the Hyades}
	\label{tab: Catalogs of the Hyades}
	\begin{tabular}{lc|lc}
		\hline\hline
		Catalog & N & Catalog & N \\
		\hline
		L & 108 & R & 360\\
		M & 17 & LR & 361 \\
		LM & 6 & RM & 46\\
		LRM & 168 & Sum & 1066\\
		\hline
	\end{tabular}
	\tablefoot{L: \citet{lodieu19}, R: \citet{roeser19}, M: \citet{meingast19}}
\end{table}   

\subsection{Rotation periods}
\label{sec: Rotation periods}
\citet{douglas19} provide rotation periods for Hyades members derived from \textit{K2}~Campaign~13 light curves. Additionally, they adopt rotation periods from \citet{douglas16} estimated from \textit{K2}~Campaign~4 
data as well as periods from \citet{radick87, radick95}, \citet{prosser95}, \citet{hartman11}, 
\citet{delorme11}, and from data of the All Sky Automated Survey \citep[ASAS,][]{pojmanski02}. We searched 
Table~3 of \citet{douglas19} for matches with our membership list and adopted 191 proposed rotation 
periods. For 13 Hyades members not listed by \citet{douglas19}, we found a rotation period 
in \citet{lanzafame18} who analyze photometric time series from \textit{Gaia} data (released within DR2). Thus 
in total, we have rotation periods for 204 Hyades members. Since \cite{douglas19} only investigate the 
Hyades core, the great majority of the stars with periods are located in the core and only 7 of the tail 
members have known periods.

\begin{figure}[t]
	\resizebox{\hsize}{!}{\includegraphics{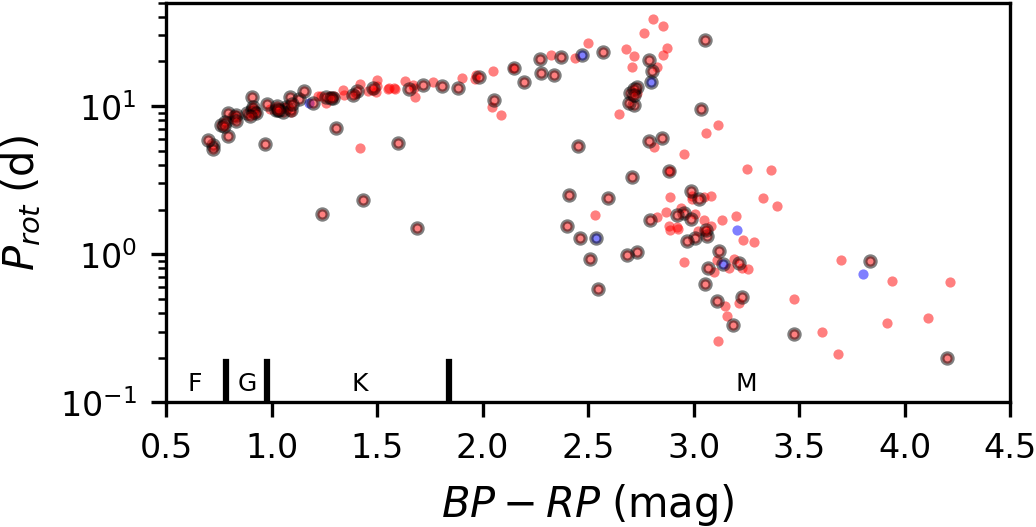}}
	\caption{Rotation periods of the Hyades as a function of the $BP-RP$ color. The red and blue sources are associated to the core and the tidal tails, respectively. Sources outlined in black are detected in X-rays.}
	\label{fig: rotation period Hyades}
\end{figure}

In Fig.~\ref{fig: rotation period Hyades} we show the rotation periods as a function of the $BP-RP$ color. 
>From F-type to early M-type stars, the sources follow the slow-rotator sequence with a steady increase of the rotation period to later spectral types. There are a few outliers with smaller periods, these sources are 
probably binaries. The rotation period abruptly drops at $BP-RP \approx 2.5$~mag near the boundary to 
fully convective stars (compare to \citet{douglas19}). The X-ray detected Hyades are marked with
black symbols in Fig.~\ref{fig: rotation period Hyades} and we discuss their properties in Sect.~\ref{sec: Rotation properties}.

\subsection{X-ray observations}
\label{sec: X-ray observations}
We adopted the X-ray properties by crossmatching the Hyades members with the second \textit{ROSAT} all-sky survey (2RXS) source catalog \citep{boller16}, the second \textit{ROSAT} PSPC catalog \citep[release 2.1.0]{2RXP-catalog} (2RXP), 
the \textit{ROSAT} HRI pointed observations \citep[3rd release]{1RXH-catalog} (1RXH), the {\it Chandra} Source Catalog \citep[specifically CSC release 2.0]{CSC-catalog}, and pointed \textit{XMM-Newton} observations that contain a Hyades member. For 
sources with multiple detections, we adopteded the X-ray data derived from \textit{XMM-Newton} data or from the CSC if available and 
otherwise the \textit{ROSAT} observation with the longest exposure time. To make the measurements of the different 
instruments comparable, we converted all X-ray fluxes into the \textit{XMM-Newton} band ($0.2-12$~keV), 
adopting an APEC thermal plasma model with a temperature of $\log(T) = 6.5$ and solar metalicity; we also estimated upper limits 
for the undetected sources using this plasma emission model. 
Note that we derived these upper limits only in a statistical way, not individually using the actual photon counts; 
for any given star, an actual
upper limit might be higher, if, for example, a nearby, source is present, or lower, if the background is exceptionally low.
Our upper limits therefore characterize the properties of our Hyades sample.  We did compare our upper limits with those given by \citet{stern95} and found in most cases our values to be rather conservative. Hence, we are confident that our upper limits are reliable for most Hyades members.

\subsubsection{\textit{ROSAT} all-sky survey}
\label{sec: ROSAT all-sky survey}
Between August 1990 and January 1991 the \textit{ROSAT} all-sky survey (RASS) was performed \citep{trumper82}. \citet{boller16} re-processed the data from this survey and created the 2RXS catalog. 
We adopted the 2RXS identifications of the Hyades members from Freund et al.~(in prep.), considering only those members that are the most likely counterpart to the 2RXS source and have an individual matching 
probability $p(H_\mathrm{lj}) > 50~\%$. 

\begin{figure}[t]
	\resizebox{\hsize}{!}{\includegraphics{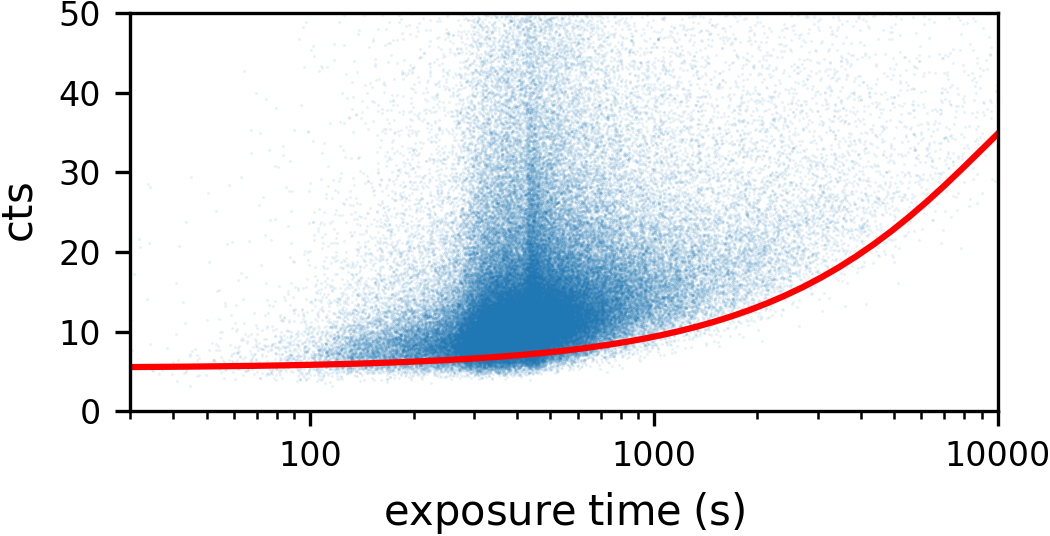}}
	\caption{Number of counts as a function of the exposure time for all 2RXS sources. The solid red curve represents the lower envelope (95~\% of the sources lie above the line).}
	\label{fig: upper limit construction 2RXS}
\end{figure}

In contrast to pointed observations, RASS covers the entire region of the Hyades, and hence, upper limits of 
the X-ray luminosity can be derived for all undetected Hyades members. To estimate the minimal number of 
counts being detectable in the RASS, we inspected the number of detected counts as a function of
exposure time for all 2RXS detections as shown in Fig.~\ref{fig: upper limit construction 2RXS}. We fitted a lower envelope to the distribution using a quadratic polynomial so that 95~\% of the sources are located 
above the envelope. Then, we applied the RASS exposure time at the position of every undetected Hyades 
member and the lower envelope to estimate the minimal number of counts that would have 
resulted in an X-ray detection, and thus, the upper limit of the countrate and the X-ray luminosity. 

\subsubsection{ROSAT pointings}
After the all-sky survey, \textit{ROSAT} performed pointed observations until the end of the mission in February 1998. 
For these observations two different instruments were available, i.e., the Position Sensitive Proportional Counter (PSPC) and the High Resolution Imager (HRI), the results are available in the 2RXP \citep{2RXP-catalog} and 1RXH \citep{1RXH-catalog} catalogs, respectively. 

We crossmatched the Hyades member list with the 2RXP and 1RXH catalogs applying a matching radius of 
$3\sigma$ of the stated X-ray positional uncertainty, excluding all 2RXP and 1RXH sources with no 
exposure time and 1RXH sources flagged as non-unique. No positional and count rate errors are given 
for $21~\%$ of the 2RXP sources detected at an off-axis angle larger than $20$~arcmin to the center of the field-of-view. For these sources, we estimated the positional uncertainty considering the width of the PSF of the Gaussian intrinsic resolution and the mirror blur\footnote{\url{https://heasarc.gsfc.nasa.gov/docs/journal/rosat_off-axis_psf4.html}} through 
\begin{equation}
\sigma = \frac{\sqrt{108.7E^{-0.888} + 1.121E^6 + 0.219\theta^{2.848}}}{\sqrt{cts}}~\mathrm{arcsec},
\label{equ: positional uncertainty 2RXP}
\end{equation}
where $cts$ is the number of source counts, $\theta$ is the off-axis angle in degrees, and for the source energy, we adopted a constant value of $E=0.2$~keV. For the 2RXP and 1RXH sources with a formal 
statistical uncertainty $<5$~arcsec, we applied a value of 5~arcsec to account for 
systematic errors. Some Hyades members are counterparts to multiple 2RXP or 1RXH sources, here, we 
calculated the weighted mean of the count rates from all 2RXP or 1RXH sources because they are likely 
detections of the same X-ray source. 

For the Hyades members covered by a 2RXP or 1RXH pointing and not detected in any of the X-ray catalogs, we 
estimated upper limits. We adopted a circular field-of-view with a diameter of $2^\circ$ and 38~arcmin for the 
2RXP and 1RXH pointings, respectively, neglecting the ribs of the PSPC and the square field-of-view of the 
HRI. We estimated the minimal number of detectable counts as a function of the exposure time in a similar 
way as in Sect.~\ref{sec: ROSAT all-sky survey}. However, since the width of the PSF strongly increases 
with the off-axis angle, we applied different lower envelops to 2RXP sources depending on their off-axis angle. We derive 2RXP upper limits only for Hyades members within 50~arcmin to the center of a 2RXP pointing 
because the number and quality of sources detected at larger angular separations is too low to fit a 
lower envelope and derive meaningful upper limits.

\subsubsection{Chandra pointings}
The {\it Chandra} X-Ray Observatory performs pointed observations since its launch in 
1999 \citep{weisskopf02}. Two different instruments, the Advanced CCD Imaging Spectrometer (ACIS) 
and the High Resolution Camera (HRC), are available for the observations. The CSC 2.0 provides information 
of about 370\,000 unique sources detected in more than 10\,000 {\it Chandra} observations with the
ACIS and HRC instruments made publicly available by the end of 2014. Pointings covering the same position were stacked 
to determine unique sources. In the region of the Hyades, \textit{Chandra} pointings cover a region of about 0.70~deg$^2$ in the core and 0.47~deg$^2$ in the tails. We searched in the CSC for counterparts to our Hyades members adopting a 
matching radius of $3\sigma$ of the {\it Chandra} positional uncertainties, applying a minimal 
uncertainty of 1~arcsec to account for systematic errors. We did not search for individual observations or 
undetected Hyades members covered by a {\it Chandra} pointing to derive upper limits. 
The CSC provides X-ray fluxes estimated from an APEC model and we adopted these values after converting 
the flux of the broad bands of the ACIS and HRC instruments to the XMM band.

\subsubsection{\textit{XMM-Newton} pointings}
We inspected all \textit{XMM-Newton} observations with a Hyades member being within 10~arcmin from the nominal aim point of the X-ray observation covering a region of about 6.7 and 1.2~deg$^2$ of the Hyades core and tails, respectively. If the source falls onto one of the three X-ray detectors, we extract the source and background counts within 10 and 20~arcsec from the proper motion corrected position on the detector(s). If 
the number of counts in any of the two considered radii is above the 99~\% interval for random background fluctuations, we consider the source as potentially detected. In that case, we consider all \textit{Gaia} DR2 sources 
that fall within 20~arcsec of the Hyades member under consideration and check if any of these potential 
alternative sources provides a better match with the X-ray photon distribution, e.g., if the centroid 
matches better with any of the alternative sources in which case we manually checked that the Hyades 
member is indeed the best matching counterpart (see Schneider et al, in prep. for details).

Hyades member that are detected by \textit{XMM-Newton} are then checked against source variability during the 
observation and the source count rate is converted into flux using the encircled energy fraction, the 
effective exposure at the detector position, and assuming a plasma temperature of $\log(T) = 6.5$ as 
used for the ROSAT conversion factor.

\section{X-ray and rotational properties of the Hyades}
\label{sec: Properties of the Hyades}
\subsection{X-ray detection rates}
\label{sec: x-ray detection rates}
In Table~\ref{tab: Catalogs of the X-ray} we provide the number of Hyades detections, unique sources, and data adopted for the following analysis for the individual X-ray catalogs as well as the number of upper limits 
and number of lowest upper limits from all X-ray catalogs. In total, 281 Hyades members are detected as
X-ray sources.
Most of the detections (212) and upper limits from pointed observations (77) are located in the core of the 
Hyades, while only 69 and 30 detections and upper limits are found in the tidal tails. The reason for the smaller number of detections in the tails is the generally larger distance of the tail 
sources. As evident in Fig.~\ref{fig: Hyades member classification}, most of the detected tail sources 
are rather close to the Sun and only a very few Hyades members are detected at larger distances. 
Furthermore, the central part of the Hyades cluster is 
much better covered by pointed X-ray observations.
\begin{table*}[t]
	\centering%
	\caption{Catalogs of the X-ray data for the Hyades}
	\label{tab: Catalogs of the X-ray}
	\begin{tabular}{lccccc}
		\hline\hline
		Catalog & Detections & Unique sources & Adopted data & Upper limits & Lowest upper limits \\
		\hline
		2RXS & 191 & 191 & 117 & 869 & 678 \\
		2RXP & 154 & 103 & 85 & 156 & 91 \\
		XMM & 63 & 58 & 58 & 4 & 4 \\
		1RXH & 44 & 39 & 19 & 22 & 10 \\
		CSC & - & 7 & 2 & 3 & 2 \\
		\hline
	\end{tabular}
\end{table*}

To further investigate the influence of the source distances on the detection fractions, we sorted the 
Hyades members into three groups according to their distance. The first group contains 65 sources closer 
than 35~pc to the Sun, with 5 sources being associated to the core and the remaining 60 to the tidal tails. 
The second group contains stars with distances between 35 and 60~pc and covers the center of the Hyades;
hence 526 of the so-selected 669 sources are classified as core sources, while 143 objects belong 
to the tidal tails.  The third group of stars has distances larger than 60~pc contains more sources in the tails (313),
while only 19 are classified as core sources. 

In Table~\ref{tab: X-ray detections Hyades distances} we provide the detection fractions of the different 
samples. As expected, the sample with the largest distance has the lowest detection fraction and the 
closest group has the highest detection fraction. Furthermore, the detection fraction depends on the 
spectral type (as adopted from the $BP-RP$ color). While the largest detection fraction are obtained for 
F- and G-type stars, the detection fraction drops for K- and M-type members due to the X-ray luminosity 
distributions of the different spectral types (cf., Sect.~\ref{sec: X-ray activity and HRD}). We do 
not find any X-ray emitting A-type Hyades members, confirming the fact that A-type stars 
are in general not strong X-ray sources  \citep{schmitt97}.
\begin{table*}[htbp]
	\centering
	\caption{X-ray detections of the Hyades with different distances to the Sun}
	\label{tab: X-ray detections Hyades distances}
	\begin{tabular}{l|lll|lll|lll}
		\hline\hline
		 & \multicolumn{3}{c|}{$d_{sun} < 35$~pc} & \multicolumn{3}{c|}{$35 < d_{sun} < 60$~pc} & \multicolumn{3}{c}{$d_{sun} > 60$~pc} \\
		SpT & all & detected & fraction $[\%]$ & all & detected & fraction $[\%]$ & all & detected & fraction $[\%]$ \\
		\hline
		A & 1 & 0 & 0 & 3 & 0 & 0 & 0 & 0 & - \\
		F & 2 & 1 & 50 & 37 & 34 & 92 & 15 & 6 & 40 \\
		G & 1 & 0 & 0 & 28 & 25 & 89 & 14 & 6 & 43 \\
		K & 14 & 7 & 50 & 96 & 38 & 40 & 46 & 3 & 7 \\
		M & 47 & 13 & 28 & 505 & 131 & 26 & 257 & 17 & 7\\
		\hline
	\end{tabular}
\end{table*}

We find 291 of the 440 sources analyzed by \citet{stern95} in our sample of Hyades dwarfs (further 34 members in \citet{stern95} have an identification in our Hyades catalog, but are not located on the main-sequence). \citet{stern95} identify 111 sources with a RASS counterpart and we add further 56 X-ray identifications of these sources. On the other hand, we do not find a reliable X-ray identification for 10 Hyades members reported by \citet{stern95} for various reasons; in some cases the angular separation between the Hyades member and the RASS counterpart is quite large and therefore we deem the identification as unreliable, or there is a better counterpart to the RASS source, either a binary companion or a background source unrelated to the Hyades cluster, and a few RASS sources found by \citet{stern95} are not part of the 2RXS catalog that we use for our study. 

\subsection{Spurious identifications}
Freund et al. (in prep.) provide matching probabilities for all 2RXS identifications of the Hyades members. Thus, 
we expect 13 of the 191 2RXS identifications to be spurious associations, however, the estimation of the matching probability does not consider that we know {\it a priori} Hyades members to be likely identifications. Hence, we expect the estimated number to be an upper bound.

To estimate the number of random 2RXP, 1RXH, XMM, and CSC associations, we randomly shifted all Hyades members several times between 3 and 10~arcmin and by a random angle and performed the same matching procedure for the shifted as for the real Hyades members. On average, we obtained 2.2 2RXP and 0.2 1RXH identifications to the shifted Hyades members and we expect a similar number of random associations for the real Hyades members. The probability of a chance alignment for XMM and CSC counterparts is thus very low (<1 spurious counterparts in our sample) according to our tests.

\subsection{X-ray activity and HRD}
\label{sec: X-ray activity and HRD}
\begin{figure*}[t]
	\sidecaption
	\includegraphics[width=12cm]{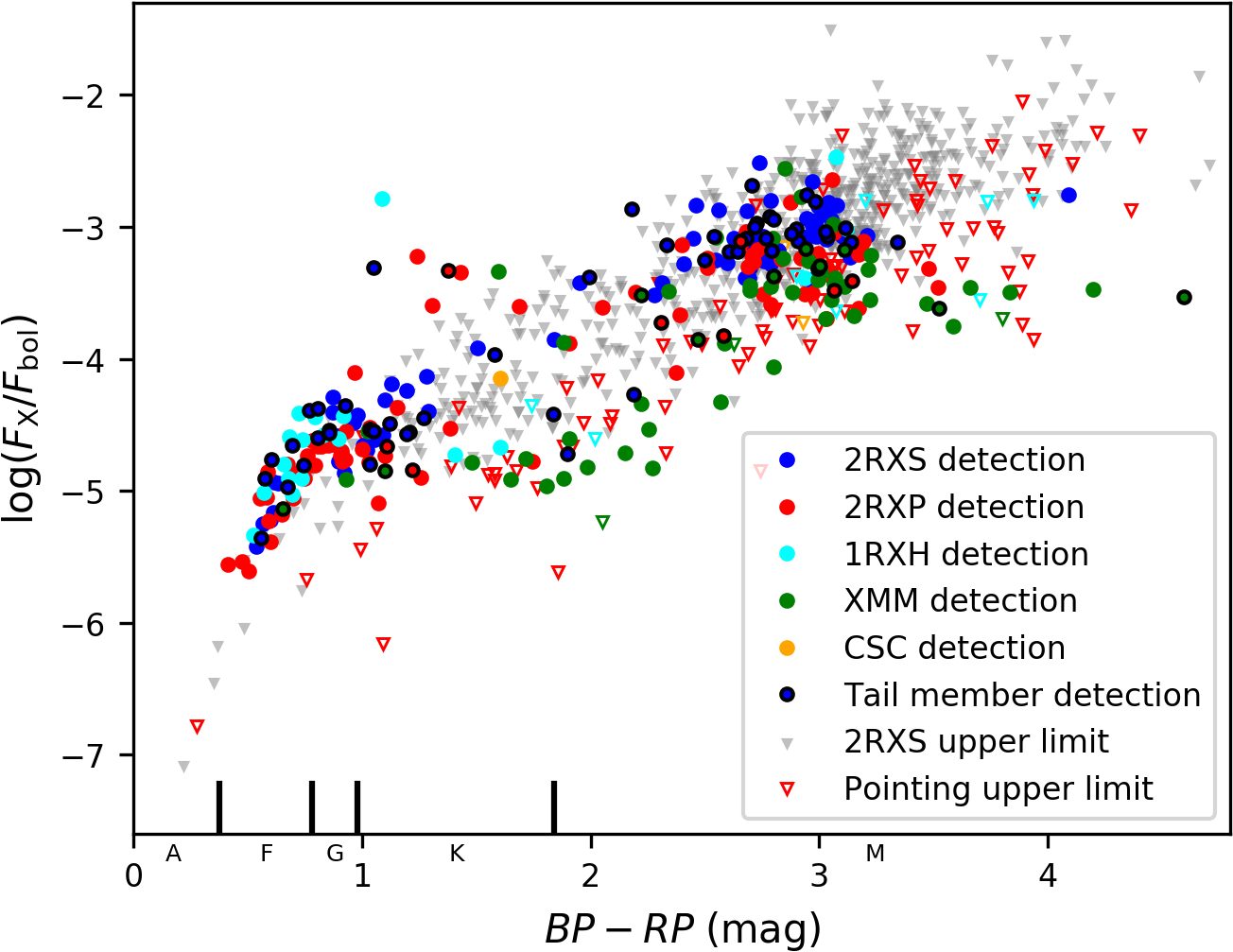}
	\caption{$F_X/F_\mathrm{bol}$ distribution of the Hyades as a function of the $BP-RP$ color. 
Blue, red, cyan, green, and orange dots indicate the values adopted from 2RXS, 2RXP, 1RXH, XMM, and 
CSC, respectively, while the triangles show the upper limits from 2RXP, 1RXH, and XMM pointings. The 
upper limits from 2RXS are shown as gray triangles in the background. In contrast to the detections 
in the core, the Hyades members in the tails are outlined in black. At the bottom we show the ranges 
of the spectral types as a guidance.}
	\label{fig: activity Hyades}
\end{figure*}

\begin{figure*}[t]
	\sidecaption
	\includegraphics[width=12cm]{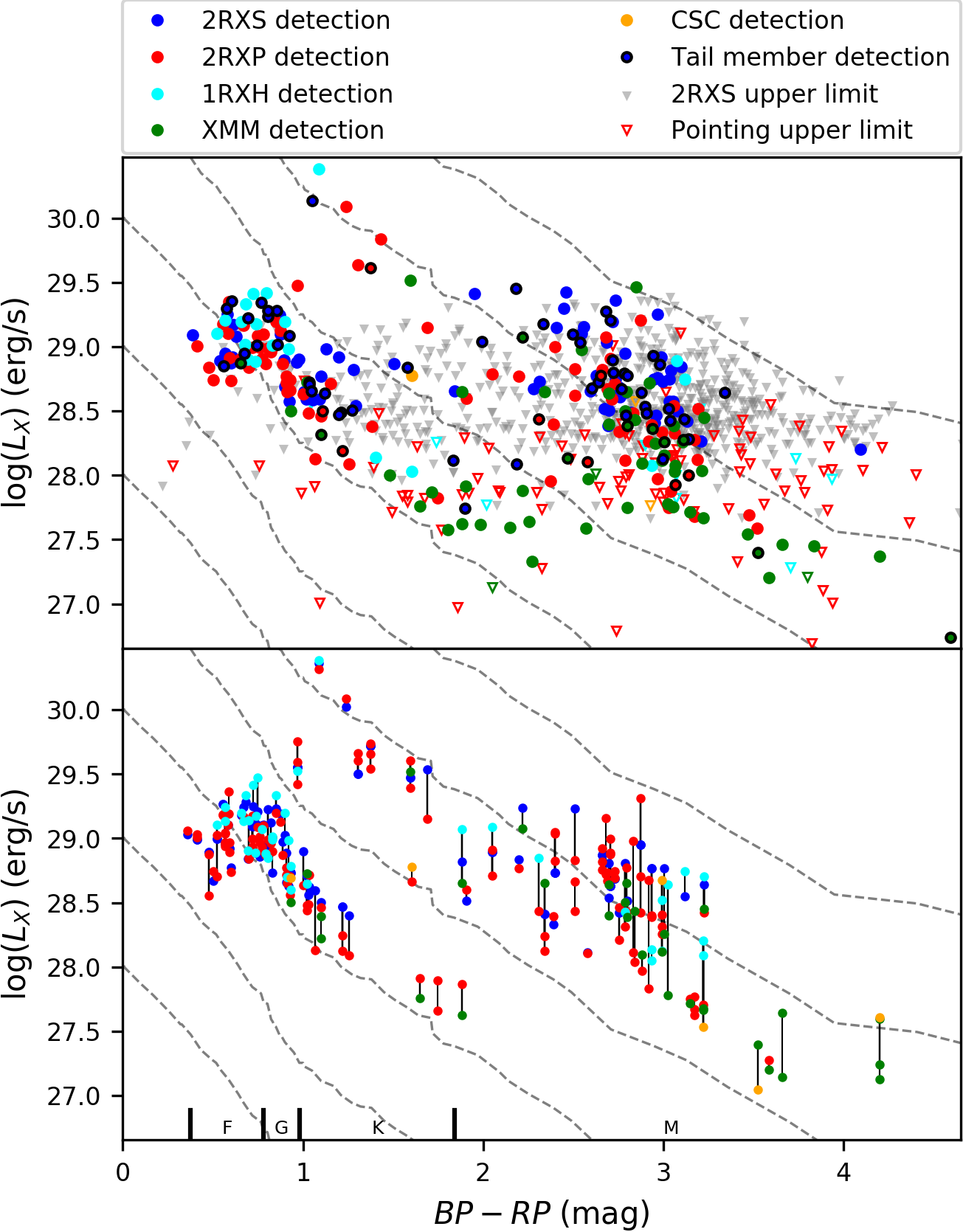}
	\caption{X-ray luminosities of the Hyades as a function of the $BP-RP$ color. The color coding is the same as in Fig.~\ref{fig: activity Hyades}. The dashed lines indicate from top to bottom activity levels of $L_X/L_{bol} =$ -2, -3, -4, -5, -6, and -7 for sources located on the main sequence. The upper panel shows the best X-ray luminosity for the individual Hyades members, while the bottom panel compares the X-ray luminosities measured by different instruments or pointings for multiple detected sources. Detections of the same source are connected by a string.}
	\label{fig: L_X HRD Hyades}
\end{figure*}

In Fig.~\ref{fig: activity Hyades} and \ref{fig: L_X HRD Hyades} we show the fractional
X-ray fluxes $F_X/F_\mathrm{bol}$\footnote{For the bolometric correction and the conversion 
between the $BP-RP$ and other photometric colors, we adopted throughout this paper the values 
given in a table based on \citet{pec13} available at \url{http://www.pas.rochester.edu/~emamajek/EEM_dwarf_UBVIJHK_colors_Teff.txt} (Version 2019.3.22)} and X-ray luminosities $L_X$ of the 
main-sequence Hyades members as a function of the $BP-RP$ color. The activity distribution of the 
F- and G-type stars is quite well defined, but the spread in fractional X-ray luminosity increases for 
later spectral types, reaching a maximum for K- and early M-type sources. Nevertheless, the X-ray 
activity (as measured by $F_X/F_\mathrm{bol}$) continuously increases towards later spectral types, 
from $\log(F_X/F_\mathrm{bol}) \approx -5.7$ for early F-type stars to the saturation limit 
at $\log(F_X/F_\mathrm{bol}) \approx -3$ for sources later 
than approximately spectral type M3  \citep{vil84, agr86, fle88, pal90}. Some sources at late spectral types
are detected above the saturation limit, which we tentatively attribute to flares.  

In Fig.~\ref{fig: L_X HRD Hyades} some early K-type sources with X-ray 
luminosities higher than $L_X = 5\times 10^{29}$~erg~s$^{-1}$ stand out particularly. These sources are X-ray bright, active binaries like 
RS CVn systems; close binaries are known to be very active in X-rays because they maintain 
their high rotation periods due to the tidal interaction \citep{wal78, dem93, dem97}.
On the other hand, Fig.~\ref{fig: L_X HRD Hyades} includes three upper limits with 
$L_X \lesssim 10^{27}$~erg~s$^{-1}$, which would be highly surprising given the age of the Hyades members.
We inspected these sources in detail and find that one data point
(Gaia DR2 3312602348628348032 at $BP-RP = 1.1$~mag)
is probably erroneously an upper limit as a likely counterpart is just outside the positional 
error region (reasonable given the number of X-ray detections). The two 
other low upper limits (at $BP-RP = 1.9$ and $2.7$~mag) are likely indicative of the fact that 
these sources are not genuine Hyades members (specifically Gaia DR2 3189577958236917632 
and 45770783575075968), which we regard compatible with the estimated contamination fraction 
in the membership list of several percent.

After excluding these three upper limits, the faintest Hyades members detected with \textit{XMM-Newton} 
have similar X-ray luminosities as the lowest upper limits. Hence, we conclude that 
all main-sequence Hyades stars with a convection zone can be detected in X-rays at sensitivities 
reachable with \textit{XMM-Newton} maybe with the only exception of the latest M-type stars.

\begin{figure*}[t]
\sidecaption
\includegraphics[width=12cm]{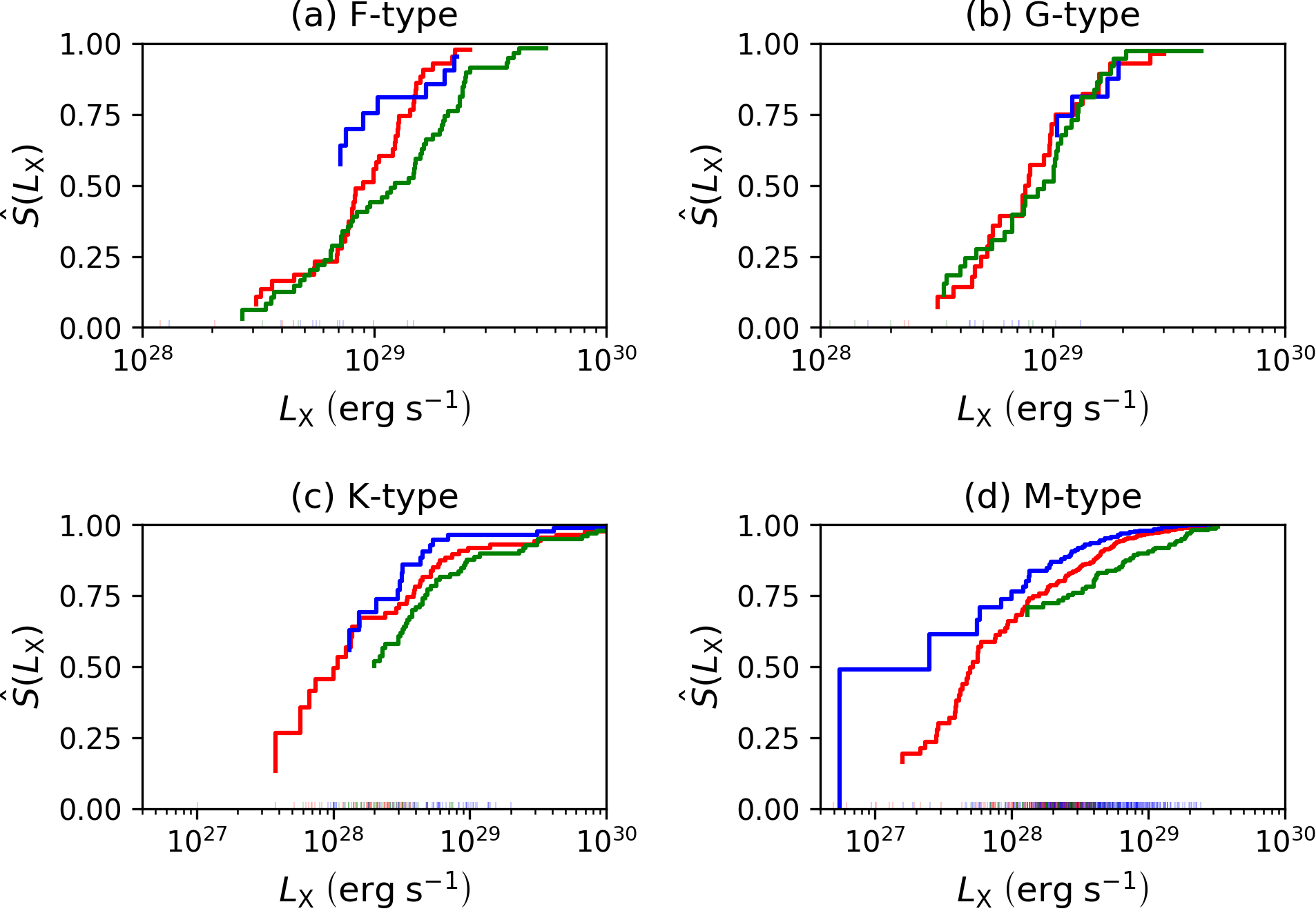}
\caption{Kaplan-Meier estimator of the X-ray luminosities for the Hyades members at different spectral types. The red, blue, and green lines show the survival functions for the core and tail members and for the sample from \citet{stern95}, respectively. The colored bars at the x-axis indicate the upper limits}
\label{fig: Hyades L_X distribution Kaplan-Meier M_G}
\end{figure*}

In Fig.~\ref{fig: Hyades L_X distribution Kaplan-Meier M_G} we compare the X-ray luminosity distributions of the core and tail members of the Hyades for different spectral types and show the distributions resulting from the data by \citet{stern95} as 
comparison; we used the Kaplan-Meier estimator to include upper limits, which is of particular importance because the detection limits of the tail sources are higher than in core due to their generally larger distances.
As visible in Fig.~\ref{fig: Hyades L_X distribution Kaplan-Meier M_G}, the earlier type stars have a steeper X-ray luminosity distribution, note the different scaling of the x-axis in the upper panels. For most spectral types, the core members are slightly brighter than the members in the Hyades tails. According to a logrank test, the X-ray luminosity distributions of the core and tail members differ with a high significance. 
However, this does not necessarily mean that the X-ray properties of the Hyades core and tail members differ intrinsically, instead, the difference is probably caused by a larger fraction of field stars spuriously associated with the Hyades tidal tails as discussed by \citet{roeser19}. Even our core members are fainter than the sample presented by \citet{stern95}. The deviations are mainly caused by the different Hyades membership lists, e.g. 4 of the 5 brightest F-type X-ray emitters in \citet{stern95} are not associated to the Hyades according to the \textit{Gaia} DR2 data that we use.  However, due to the detections from pointed observations especially with
XMM-Newton, we can extend the X-ray luminosity distribution to fainter luminosities, which is particularly important for the K- and M-type Hyades members.

\subsection{Variability}
\label{sec: Variability}
\begin{figure}[t]
	\resizebox{\hsize}{!}{\includegraphics{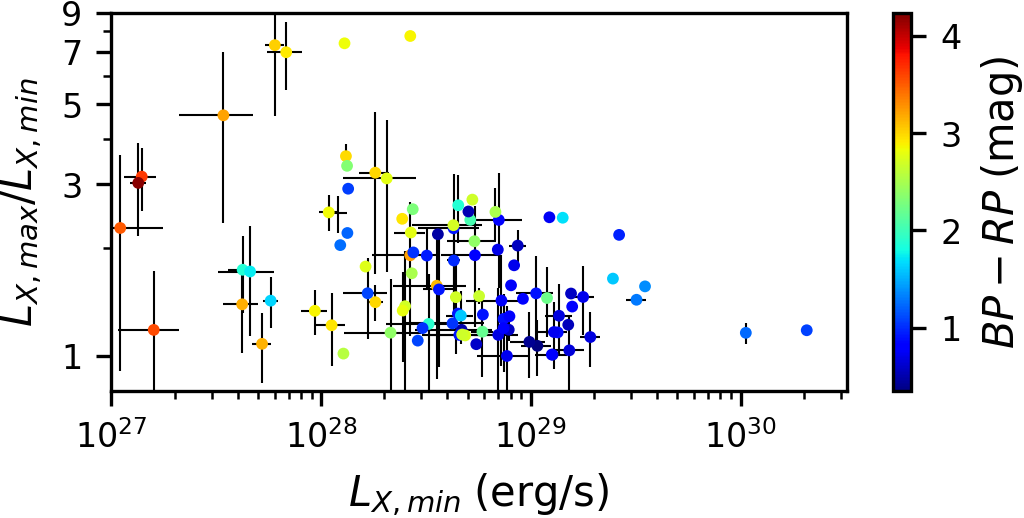}}
	\caption{Fraction of the maximal to minimal measured X-ray luminosity for multiple detected Hyades members. The color scales with the $BP-RP$ color. No error is given for 2RXP sources detected at an off-axis angle larger than 20~arcmin to the center of the pointing. }
	\label{fig: L_X comparision Hyades}
\end{figure}

X-ray luminosities from different epochs are available for 104 Hyades members detected by different 
instruments or by the same instrument in different pointings. In the bottom panel of 
Fig.~\ref{fig: L_X HRD Hyades} and in Fig.~\ref{fig: L_X comparision Hyades}, we compare the X-ray 
luminosities of the various detections, we do not include upper limits because our upper limits are suited to describe the properties of samples but they might be unreliable for individual sources. Although the observations are up to 25 years apart, in most 
cases the X-ray luminosities measured for each source do not differ by more than a factor of three. Hence 
we conclude that the Hyades members are not strongly affected by activity cycles compared to the Sun,
whose X-ray luminosity differs by more than one order of magnitude between solar minimum and 
maximum \citep{peres00} in a comparable energy band. 

Hyades members that show larger differences between epochs are particularly red and, thus, 
low-mass stars. These stars are generally more active and therefore show more frequent flaring.
Such stochastic flares can increase the X-ray flux by a factor of more than 100 \citep[e.g.][]{stelzer06}. 
Hence, the larger differences in the X-ray fluxes derived at two epochs 
are likely caused by different degrees of flaring activity.

This agrees well with the results reported by \citet{stern95}, who find most of the Hyades members to vary by not more than a factor of 2 and attributed most of the larger variations to flares. Similarly, \citet{micela96} find a variability by a factor of 2 for only about 15~\% of the much younger Pleiades members; thus X-ray flaring does not corrupt these cross-mission comparisons.

\subsection{Rotation properties}
\label{sec: Rotation properties}
Out of 204 Hyades members with measured rotation periods, 103 are detected in X-rays. As shown in 
Fig.~\ref{fig: rotation period Hyades}, the detection fraction generally decreases with increasing 
rotation period. Hence, most of the potential binaries located below the slow-rotator sequence and 
many of the earlier fully convective stars are detected in X-rays. However, despite their short 
rotation periods, the detection fraction decreases for the latest spectral types because of their 
low bolometric luminosities.

\begin{figure}[t]
	\resizebox{\hsize}{!}{\includegraphics{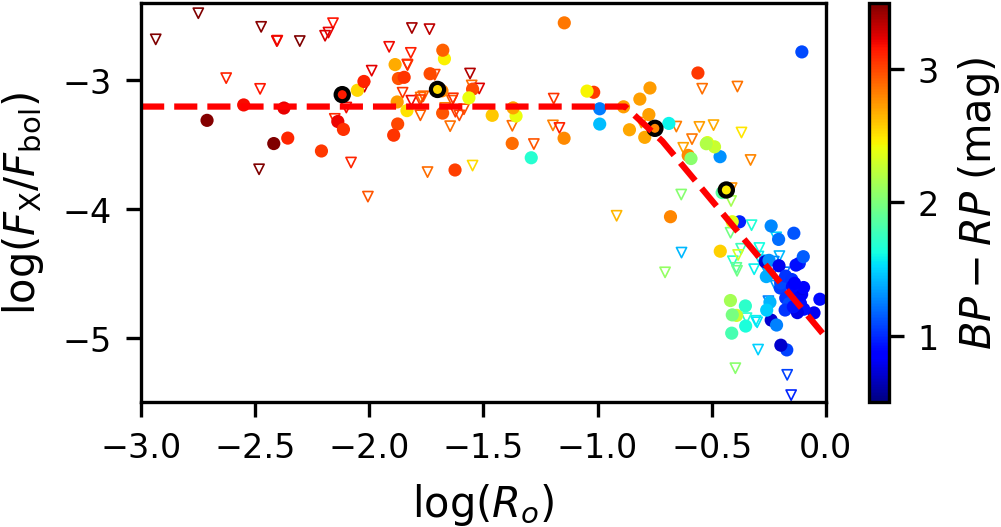}}
	\caption{X-ray activity as a function of the Rossby number for Hyades members with known 
rotation period. The dots indicate sources detected in X-rays, while the triangles show upper limits. 
In contrast to the sources in the core, the Hyades members in the tails are outlined in black. 
The color scales with the $BP-RP$ color. The red dashed line indicates the best fit.}
	\label{fig: activity vs rossby from color}
\end{figure}

For the Hyades members with known rotation periods, we estimated the Rossby number adopting 
empirical convective turnover times provided by Equation~5 in \citet{wright18}, which is  an improved version 
of Equation~10 from \citet{wright11}. However, for 4 Hyades members with known rotation periods, the 
color is outside the range in which the correlation from \citet{wright18} is valid. In 
Fig.~\ref{fig: activity vs rossby from color} the X-ray activity as measured by $F_X/F_\mathrm{bol}$
is plotted as a function of the Rossby number. Most sources with $\log(R_o) \lesssim -1.0$ are 
saturated, these sources are generally the later type stars. The activity linearly decreases for 
larger Rossby numbers and stars of earlier spectral types also have larger Rossby numbers;
we note that the outlier,
which is saturated but has a small Rossby number is a binary of Algol type. 

Following \citet{pizzo03, wright11, wright18}, we fit the relationship between the activity indicator
$R_X =\log(F_X/F_\mathrm{bol})$ and the Rossby number $R_o$ by the ansatz
\begin{equation}
R_X = 
\begin{cases}
R_{X,\mathrm{sat}} & \log(R_o) < \log(R_{o,\mathrm{sat}}) \\ 
\log(R_o)\cdot\beta + C & \log(R_o) \geq \log(R_{o,\mathrm{sat}}). \\
\end{cases}
\end{equation}
where $R_{X,\mathrm{sat}}$ is the saturation limit, $R_{o,\mathrm{sat}}$ is the Rossby number at the saturation threshold to the saturation regime, $\beta$ is the power-law index, and the constant is given by $C = R_{X,\mathrm{sat}} - \log(R_{o,\mathrm{sat}})\cdot \beta$. 
We applied the method of least squares to the detections in Fig.~\ref{fig: activity vs rossby from color} 
excluding the outlier and find a saturation limt of $R_{X,\mathrm{sat}} = -3.21 \pm 0.16$, a break value 
between saturated and unsaturated sources at $\log(R_{o,\mathrm{sat}}) = -0.84 \pm 0.18$, and a power 
law index of $\beta = -2.12 \pm 0.60$. Unlike \citet{wright11}, we did not explicitly exclude flares from our sample, nevertheless, the break value to the saturation regime and saturation limit of the X-ray activity from our Hyades sample is very similar to the values obtained by \citet{wright11}. Therefore, we conclude again that the 
X-ray detections of the Hyades members are not strongly biased by flares. We find a power-law index 
that is slightly steeper than the canonical value of $\beta = -2$ but the canonical value and the value from \citet{wright11} are within the error margins.

\section{Conclusion}
\label{sec: Conclusion}
In this paper we present an update on the X-ray and rotational properties of the Hyades. 
Using membership criteria based upon the recent \textit{Gaia} DR2 release, we create a membership list
containing 1066 main-sequence Hyades stars both in the core and in the tidal tails of the
Hyades cluster.  Using all available X-ray data, we detect 281 Hyades members ($\sim$ 26\%)
as X-ray sources, which is a significant increase compared to 139 unique Hyades detections 
reported by \citet{stelzer01} and 187 unique Hyades members reported by \citet{stern95}; note that 
due to differences in the Hyades membership list, the RASS data reduction, and the identification 
procedure, we have only 111 sources in common with \citet{stern95}. 

Confirming earlier results from \citet{micela88} and \citet{stern95}, we find the highest detection fractions for 
F- and G-type stars, while the detection fraction decreases for K- and M-type members. 
We specifically detect Hyades members with X-ray luminosities in the range from 
$\sim 2\times 10^{27}$~erg~s$^{-1}$ to $\sim 2\times 10^{30}$~erg~s$^{-1}$ covering three 
orders of magnitude. The brightest sources are active binaries, while late M-type dwarfs reach the 
lowest X-ray luminosities. The X-ray luminosity distributions of F- and G-type members are much steeper than for K- and M-type dwarfs. Compared to \citet{stern95}, we extend the distributions of the latest spectral types to lower luminosities. Since we do not find meaningful upper limits below 
$1\times 10^{27}$~erg~s$^{-1}$, we expect that all Hyades members with a convection zone are actually 
detectable with sufficiently long \textit{XMM-Newton} exposures. Furthermore, the upcoming \textit{eROSITA} 
all-sky survey \citep[eRASS]{eROSITA} is expected to provide nearly complete X-ray detections for the 
Hyades members, at least in the core. The observed X-ray activity ranges from 
$\log(F_X/F_\mathrm{bol}) \approx -5.7$ for early F-type sources and continuously increases for 
decreasing spectral type reaching the saturation limit at $\log(F_X/F_\mathrm{bol}) \approx -3$ for 
sources later than spectral type M3. F- and G-type stars show the smallest spread in their 
X-ray luminosities, however, the intrinsic spread increases for K- and M-type sources just like
the spread in the observed rotation periods.
For 104 Hyades members, we find multiple detections  either by different instruments or in 
different pointings. Similar to \citet{stern95} who compare X-ray luminosities from \textit{Einstein} 
and RASS, we do not find strong variations attributable to solar-like activity cycles, although the 
observations have different time intervals with a maximum baseline of 25 years. Only a few Hyades members show variations larger than a factor of three and these sources are all M-type dwarfs, where flares are probably responsible for the variation. 

For the first time, we analyze the X-ray properties of the Hyades tidal tails. The X-ray detection 
fraction is lower in the tails because of generally larger source distances. The X-ray luminosity distributions 
formally differ between the core and tails of the Hyades, however, this difference does not necessarily argue 
against their common origin but is likely caused by a higher contamination of field stars in the tail member sample.
The now running \textit{eROSITA} all-sky survey is expected to provide far more detections among the
tidal tail members and will therefore allow a far more detailed comparison between core and tails.

We find rotation periods for 204 of our bona fide Hyades members, 103 of them ($\sim$ 50\%)
are detected in X-rays, and Rossby numbers can be estimated.  In an activity-rotation diagram,
the linear increase of activity with decreasing Rossby number for slow rotators is visible as well 
as the saturation limit for the fast rotators. For the Hyades, sources with high and low Rossby  
numbers differ mainly by spectral type, however, the parameters of the activity-rotation-relation are 
very similar to those obtained by \citet{wright11,wright18} who applied sources of different 
ages. The ongoing Transiting Exoplanet Survey Satellite (TESS) mission will significantly improve on the
number of available rotation  periods for Hyades members and thus again allow for a more detailed
comparison between core and tails.

\begin{acknowledgements}
	SF acknowledge supports through the Integrationsamt Hildesheim, the ZAV of Bundesagentur f\"ur Arbeit, and the Hamburg University, JR by DLR under grant 50 QR 1605 and PCS through the SFB 676 founded by DFG and by DLR under grant 50 OR 1901. SF thanks Gabriele Uth and Maria Theresa Sangdaan-Lehmann for their support. This work has made use of data from the European Space Agency (ESA) mission
	{\it Gaia} (\url{https://www.cosmos.esa.int/gaia}), processed by the {\it Gaia}
	Data Processing and Analysis Consortium (DPAC,
	\url{https://www.cosmos.esa.int/web/gaia/dpac/consortium}). Funding for the DPAC
	has been provided by national institutions, in particular the institutions
	participating in the {\it Gaia} Multilateral Agreement. The paper is based on observations obtained with \textit{XMM-Newton}, an ESA science mission with instruments and contributions directly funded by ESA Member States and NASA. This research has made use of data obtained from the Chandra Source Catalog, provided by the Chandra X-ray Center (CXC) as part of the Chandra Data Archive. This paper includes data collected by the \textit{K2} mission. Funding for the \textit{K2} mission was provided by the NASA Science Mission directorate. We used the VizieR catalog access tool and the SIMBAD database, operated at CDS, Strasbourg, France. The original description of the VizieR service was published in A\&AS 143, 23. 
\end{acknowledgements}

\bibliographystyle{aa} 
\bibliography{mybib}

\end{document}